\documentclass[times,twocolumn,final]{elsarticle}

\usepackage{cag}
\usepackage{framed,multirow}
\usepackage{tabularx}
\usepackage{graphicx}

\usepackage{amssymb}
\usepackage{latexsym}

\usepackage{url}
\usepackage{xcolor}
\definecolor{newcolor}{rgb}{.8,.349,.1}

\usepackage{hyperref}

\usepackage[switch,pagewise]{lineno} 

\journal{Computers \& Graphics}

\begin{document}

\verso{Preprint}

\begin{frontmatter}

\title{Swarm manipulation: An efficient and accurate technique for multi-object manipulation in virtual reality}%

\author[1]{Xiang Li}
\cortext[cor1]{Corresponding author at: Department of Engineering, University of Cambridge, Cambridge, Cambridgeshire, United Kingdom.}
\emailauthor{xl529@cam.ac.uk}{\bf X. Li}

\author[2]{Jin-Du Wang}

\author[1]{John J. Dudley}
\author[1]{Per Ola Kristensson}

\address[1]{Department of Engineering, University of Cambridge, Cambridge, Cambridgeshire, United Kingdom}
\address[2]{School of Software Engineering, Xi'an Jiaotong University, Xi'an, Shaanxi, China}

\received{\today}

\begin{abstract}

The theory of swarm control shows promise for controlling multiple objects, however, scalability is hindered by cost constraints, such as hardware and infrastructure. Virtual Reality (VR) can overcome these limitations, but research on swarm interaction in VR is limited. This paper introduces a novel Swarm Manipulation interaction technique and compares it with two baseline techniques: Virtual Hand and Controller (ray-casting). We evaluated these techniques in a user study ($N$ = 12) in three tasks (selection, rotation, and resizing) across five conditions. Our results indicate that Swarm Manipulation yielded superior performance, with significantly faster speeds in most conditions across the three tasks. It notably reduced resizing size deviations but introduced a trade-off between speed and accuracy in the rotation task. Additionally, we conducted a follow-up user study ($N$ = 6) using Swarm Manipulation in two complex VR scenarios and obtained insights through semi-structured interviews, shedding light on optimized swarm control mechanisms and perceptual changes induced by this interaction paradigm. These results demonstrate the potential of the Swarm Manipulation technique to enhance the usability and user experience in VR compared to conventional manipulation techniques. In future studies, we aim to understand and improve swarm interaction via internal swarm particle cooperation.
\end{abstract}


\end{frontmatter}


\section{Introduction}

A \textit{swarm} is a concept inspired by natural phenomena where groups of individual entities operate collectively to accomplish tasks \cite{9460560}. Examples in nature include bird flocking, ant foraging, and fish schooling, where simple entities form complex behaviors through local interactions \cite{sneyd2001self}. Typically, swarm interaction involves extracting engineering principles from the study of natural systems to enable comparable capabilities for multi-agent systems (e.g., multi-robot systems) with similar abilities~\cite{bonabeau_inspiration_2000}.

The realm of robotics and drones has seen the effective implementation of swarm control theory~\cite{saffre_design_2021,haghighi_robotic_2016,yu_aerorigui_2023,kolling_human_2016}. Swarm control enables the coordination of numerous entities as a collective to achieve a common goal, promoting efficient simultaneous selection and manipulation of multiple objects~\cite{nakagaki_disappearables_2022,nakagaki_hermits_2020,suzuki_reactile_2018}. The benefits are substantial, including scalability, fault tolerance, and robustness~\cite{chen_toward_2020}, which allows systems to adapt to varying numbers of entities, while fault tolerance ensures system functionality despite individual agent failures~\cite{brown_human-swarm_2014}. Swarm systems also exhibit dynamic adaptability, adjusting collectively to evolving conditions~\cite{chen_toward_2020}. Despite these advantages, there are hindrances to the widespread implementation of swarm control, such as high costs associated with hardware, communication infrastructure, and computational resources, as well as the need for significant maintenance and expertise~\cite{Lars_2023}. However, within the virtual reality (VR) environment, many of these limitations of swarm control implementation can potentially be mitigated. This makes the prospect of swarm interactions in VR a compelling area of exploration, with the potential to allow us to reimagine the way we interact with virtual objects~\cite{khodr_being_2020}. 

\begin{figure*}[t]
  \centering
  \includegraphics[width=\linewidth]{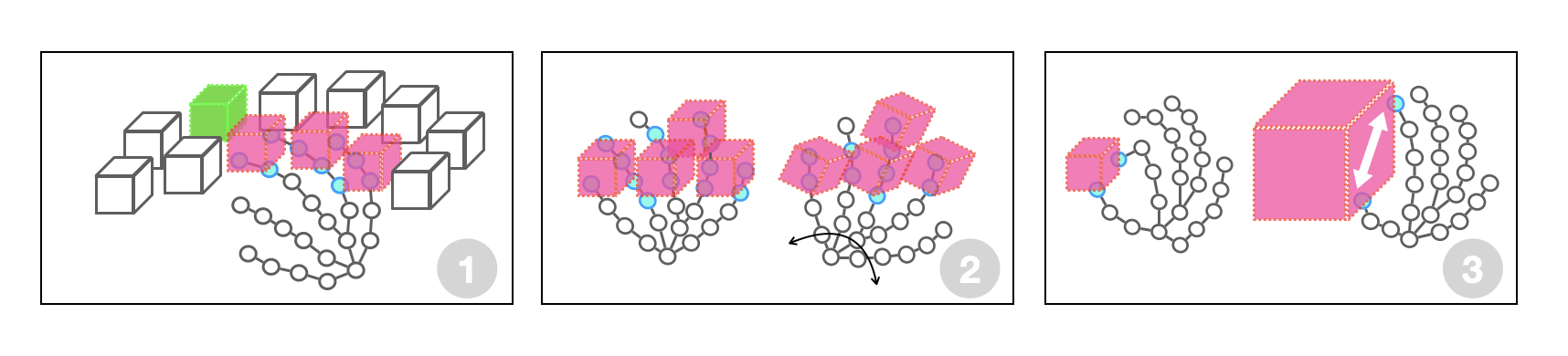}
  \caption{We introduce the Swarm Manipulation technique in Virtual Reality (VR) and compare it with two conventional manipulation techniques: Hand and Controller (Ray-Casting). We evaluate these techniques based on three tasks: (1) Selection, (2) Rotation, and (3) Resizing. During object selection, the swarm particles transition from their original color to blue.}
  \label{fig:teaser}
\end{figure*}

By leveraging virtual environments, VR can simulate large-scale swarm interactions without the need for extensive physical infrastructure. This not only reduces costs but also offers the flexibility to experiment with swarm configurations and behaviors in a controlled setting. 
The notion of swarm interaction offers a new lens through which human perception can be augmented in virtual environments. 
Swarm interaction enables a user to control and influence a collective of virtual swarm agents.
This paper focuses on studying techniques that enable user control of virtual swarm agents.
We refer to this task of controlling the swarm as swarm manipulation.

Exploring novel manipulation techniques and providing a simulated environment for intuitive and immersive interactions with virtual objects has emerged as a promising direction for improving user experience in VR~\cite{bergstrom2021evaluate}. Traditional techniques of object selection and manipulation in VR rely on avatar hand movements, which mimic the movements of the user's real hands, or controller-based ray-casting for object selection and allow for adjustment of movement gain to reach distant objects~\cite{mendes_survey_2019,hinckley_survey_1994,bowman_evaluation_1997}. For instance, users have the flexibility to utilize both hand controllers for near and far field interactions on the Quest. Similarly, HoloLens incorporates a ray-casting technique from the hand to facilitate far-field interaction. While these methods are practical, there is a trade-off between intuitiveness and efficiency. An emergent manipulation technique is \textit{Ninja Hand}~\cite{schjerlund_ninja_2021}, where the user's real hands in VR are mapped to multiple virtual hands. This approach holds promise in reducing time spent moving and decreasing overall workload. However, it comes with its own challenges, such as causing visual clutter due to the presence of numerous hands in the virtual space, which can limit and even reverse its benefits with increasing hand count~\cite{schjerlund_ninja_2021}.

This paper explores an alternative way of disassociating the virtual from the physical by exploring a novel interaction technique (see Figure \ref{fig:teaser}) to bestow users with the ability to select and manipulate multiple virtual objects simultaneously. In this paper, we define \emph{swarm manipulation} as a method that leverages a group of swarm particles, where each particle operates individually but exhibits collective behaviour in response to human actions, enabling the simultaneous selection and manipulation of multiple objects. To realize this idea, we first present the results of a user study ($N = 12$) that aims to provide us with an indication of the effectiveness of a Swarm Manipulation technique compared to other popular manipulation techniques in VR. This study focuses on evaluating three manipulation techniques: Virtual Hand, Controller (ray-casting), and Swarm Manipulation, which are assessed in three tasks: selection, rotation, and resizing, across five conditions: single-target close-distance, single-target long-distance, dual-target close-distance, dual-target long-distance, and all-target. We then present the results of a second user study ($N = 6$). This study builds upon our prior research~\cite{li2023swarm} by providing new empirical findings and exploring the potential future applications of swarm manipulation in VR.

In summary, the contributions of this paper can be summarized as follows:

\begin{itemize}
    \item To our knowledge, this is the first exploratory user study using the proof-of-concept swarm interaction in VR. We provide an in-depth analysis of the effectiveness and user experience of the Swarm Manipulation technique compared to the other two baseline techniques in VR: Hand and Controller.
    \item We investigate the advantages and limitations of the Swarm Manipulation technique in terms of task performance (i.e., the task completion time and deviation), perceived workload, usability, and novelty.
    \item We collect users' subjective opinions in open-ended VR scenarios and discuss how the ownership of discrete forms (swarm-like) of the body changes.
    \item Our study results offer valuable insights for developers and designers seeking to integrate Swarm Manipulation techniques into future VR applications. Furthermore, we envision the potential future developments and possibilities of swarm interactions.
\end{itemize}

\section{Related Work}
In this section, we reviewed the existing literature and research pertaining to object manipulation in VR and swarm interaction systems. The goal is to identify the gaps and opportunities for integrating swarm interactions into manipulation tasks in VR.

\subsection{Object Manipulation in VR}
Object manipulation in VR involves interacting with and manipulating virtual objects within a simulated environment~\cite{veit_influence_2009,mendes_benefits_2016,dudley2018bare}. Numerous techniques and approaches have been explored to enhance the object manipulation experience in VR \cite{hinckley_survey_1994,mendes_survey_2019,difeng_manipulation,difeng_onbody}. For example, Virtual Hand, a widely used mid-air interaction paradigm in modern VR systems, allowed users to manipulate objects in virtual environments~\cite{mendes_survey_2019,xu_exploring_2020}. To enhance this interaction technique, several approaches had been developed. One such approach was \textit{Go-Go}~\cite{poupyrev_go-go_1996} and its extensions~\cite{mine1997moving,forlines2006hybridpointing}, which enabled users to reach distant targets by using a non-linear mapping between the controlled motion and the effected motion~\cite{bowman_evaluation_1997}. Additionally, techniques like Ray-Casting and scaling down the virtual world had been employed to interact with out-of-reach objects~\cite{pierce_voodoo_1999,stoakley_virtual_1995,grossman_06}. Dewez et al. developed "avatar-friendly" manipulation techniques~\cite{dewez_influence_2019} to strengthen the bond between users and their virtual avatars. The researchers argued that a more intuitive and efficient way of manipulating objects in VR could lead to more immersive experiences. A decade earlier, Slater et al.~\cite{slater_first_2010} conducted a study on the body transfer phenomenon in VR, demonstrating that if users could manipulate a virtual body as their own, it significantly enhanced the sense of presence and embodiment in VR. This research highlighted the critical role of efficient object manipulation in creating immersive VR experiences.

\subsection{Swarm Interactions}
Swarm interactions involves studying human-swarm interaction (HSI) and identifying fundamental principles and invariants. Brown et al.~\cite{brown_two_2016} proposed two invariants for geometric-based swarms: the collective state and the balance between span and persistence. Brown et al.~\cite{brown_human-swarm_2014} emphasized managing attractors to individual abstract agents and focused on collective behavior. Kolling et al.~\cite{kolling_human_2016} surveyed human-swarm interaction, while Kolling et al.~\cite{kolling_human-swarm_2013} compared intermittent and environmental interaction types. Dietz et al.~\cite{dietz_human_2017} explored the human perception of swarm robot motion.

Swarm user interfaces also introduced innovative concepts and architectures for interactive systems. Nakagaki et al. presented \textit{HERMITS}~\cite{nakagaki_hermits_2020} and \textit{(Dis)Appearables}~\cite{nakagaki_disappearables_2022}, an architecture that enabled dynamic reconfiguration of self-propelled Tangible User Interfaces (TUIs) using mechanical shell add-ons and actuated these swarm TUIs to appear and disappear dynamically. Yu et al.~\cite{yu_aerorigui_2023} introduced \textit{AeroRigUI}, an actuated TUI for 3D spatial interaction using controlled strings attached to ceiling surfaces. Le Goc et al.~\cite{le_goc_zooids_2016} presented \textit{Zooids}, an open-source platform for swarm user interfaces, while Suzuki et al.~\cite{suzuki_reactile_2018} explored \textit{Reactile}, an approach to programming swarm user interfaces through direct physical manipulation. These studies offered valuable insights into the potential and design considerations of swarm user interfaces, contributing to advancements in interactive systems and human-robot interaction.

\subsection{Mapping from One to Many}
Swarm manipulation can also be considered as a mapping mechanism from one to many in VR, which has been explored in previous research. Some studies suggested using multiple limbs or additional fingers to enhance the VR experience, focusing on aspects like body acceptance and ownership~\cite{hoyet2016wow}. However, it remained unclear whether these additions improved performance in interactive tasks. \textit{Ninja Cursors}~\cite{10.1145/1357054.1357201} addressed this issue by mapping input from a single mouse to multiple virtual cursors distributed across a desktop display. This approach improved target acquisition efficiency for large 2D displays. Only one cursor could actively hover over a target at a time, while the others waited in a queue. Gaze tracking had also been incorporated into this technique, as seen in the rake cursor and the work by Räihä and Špakov~\cite{raiha2009disambiguating}, to choose which cursor was active. Lubos et al.~\cite{lubos20144} used head tracking in VR to disambiguate between two sets of virtual hands but did not investigate the impact of manipulating the number of hands on shortest-distance gains. \textit{Ninja Hand}~\cite{schjerlund_ninja_2021} mapped the user's real hand to multiple virtual hands in VR, and this approach held promise in terms of reduced movement times and lower overall workload. 

However, each of these methods visually altered the basic shape of the body, and the presence of a large number of cursors or hands required users to adapt and learn how to use these techniques, resulting in an increased workload~\cite{schjerlund_ninja_2021,li2021}. This could potentially make these methods less user-friendly and more challenging to use efficiently.

\section{Swarm Manipulation}

\begin{figure}[t]
  \centering
  \includegraphics[width=\linewidth]{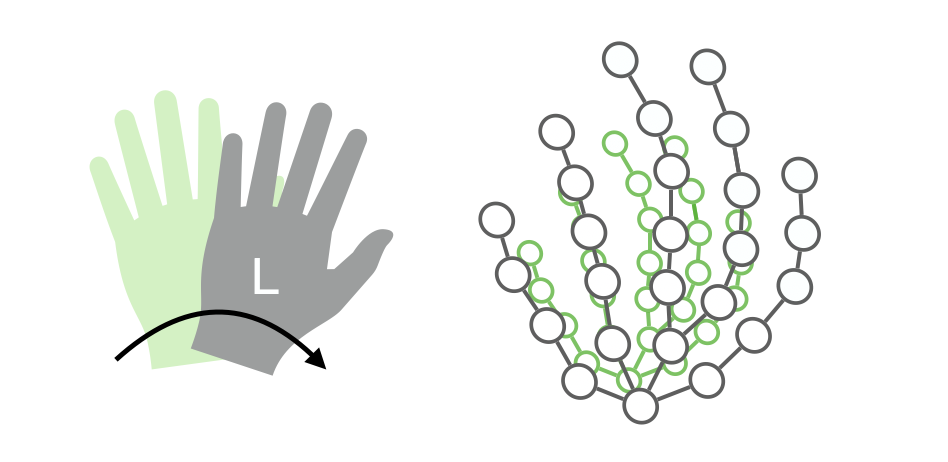}
  \caption{The figure illustrates the Swarm Hand concept consisting of two key components. The first component, the Swarm Hand, is composed of a swarm of particles, controlled by users through hand gestures. The second component, the non-dominant hand depicted as the left hand (L) in the figure, enables adjustment of distribution levels. This feature determines the size of the Swarm Hand and allows users to grasp or select multiple virtual objects based on their preferences.}
  \label{fig:distribution}
\end{figure}

In this paper, we present \emph{Swarm Manipulation}, a novel interaction technique for object manipulation in virtual environments. This technique consists of two main components, each serving a specific purpose. The first component is the dominant Swarm Hand, represented by the right ``hand'' in Figure \ref{fig:distribution}. This hand is composed of a swarm of particles that users can control through discrete hand movements. We utilize Meta Quest’s built-in hand tracking detection to identify hand joints, focusing specifically on the fingertip and palm joints. We generate swarm particles at intermediate points along the connections between each fingertip and the palm, allowing the user to control these particles through hand movements to select and manipulate virtual objects. The Swarm Hand incorporates the Go-Go technique~\cite{poupyrev_go-go_1996}, which enables non-linear mapping between the user's hand movements and the behaviour of the swarm. This technique allows for intuitive and flexible manipulation of virtual objects within the VR environment. By leveraging discrete hand movements, users can easily navigate and interact with the swarm to perform various tasks and actions. 

The second component of the Swarm Manipulation technique is the non-dominant hand, depicted by the left hand in Figure \ref{fig:distribution}. This hand plays a crucial role in adjusting the level of distribution within the swarm. The distribution level defines the volume of the swarm hand, determining how many proximate objects it can grasp. Users can modify the swarm's distribution level by rotating their wrist~\cite{song2022efficient}, providing a simple and intuitive means to control the graspability of objects within the virtual environment. To provide visual feedback and facilitate interaction, a semicircular panel surrounding the user's wrist displays the current level of swarm distribution. This visual indicator allows users to perceive and monitor the volume and graspability of the swarm hand, aiding them in making informed manipulation decisions. 

\subsection{Motivation}

Our objective is to assess the usability of a novel proof-of-concept swarm manipulation technique in VR. To achieve this, we conducted two separate user studies with the following objectives: (1) comparing this innovative technique with two prevalent interaction methods employed in VR, specifically Controllers (ray-casting) and Virtual Hands, and (2) gathering comprehensive user feedback concerning the perceptual aspects and the impact on users' sense of immersion when employing swarm manipulation within open-ended scenarios in VR.

\section{User Study 1: Comparing Swarm Manipulation with Controller and Virtual Hands in VR}

We carried out our first user study to investigate the effectiveness of the Swarm Manipulation technique compared to other popular manipulation techniques in VR. The study aimed to evaluate the performance and user experience of three manipulation techniques: Virtual Hand, Controller (ray-casting), and Swarm Manipulation. The primary objective was to examine the advantages and limitations of the Swarm Manipulation technique compared to the other techniques. This user study contributes to the existing knowledge of manipulation techniques in VR, specifically investigating the effectiveness of the Swarm Manipulation technique.

\subsection{User Study 1: Method}

\subsubsection{Study Design}

\begin{figure}[t]
  \centering
  \includegraphics[width=0.95\linewidth]{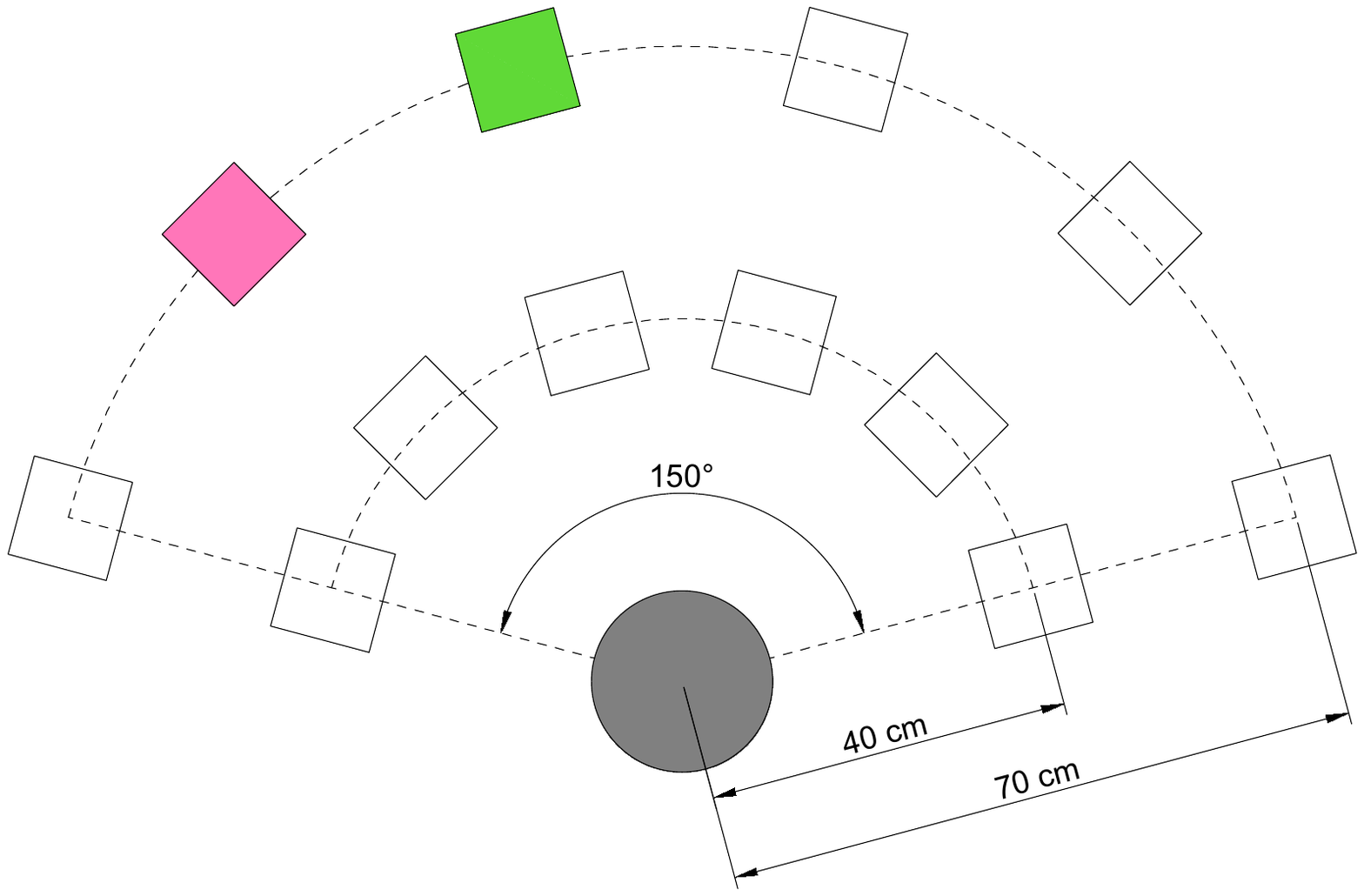}
  \caption{There are twelve objects that should be manipulated during the study. The selected objects are highlighted in red, while the target object is highlighted in green.}
  \label{object}
\end{figure}

The user study utilized a repeated measures factorial design with two independent variables: (a) \textit{Manipulation Technique} (Hand, Controller, and Swarm), and (b) \textit{Condition} (single-target close-distance, single-target long-distance, dual-target close-distance, dual-target long-distance, and all-target) (see Figure \ref{object}). The experiment consisted of three tasks: Selection, Rotation, and Resizing, as suggested by Bowman et al.~\cite{bowman1999testbed}. 

The five conditions were determined based on two main factors: Control Number (single vs. dual selection) and Distance (close-distance vs. long-distance), resulting in four combinations. This is because when selecting objects in VR, the distance of the object can be differentiated according to the state of the person's arm~\cite{xu_exploring_2020}, for example, between close range where the arm is bent (i.e., around 40 cm) and far range where the arm needs to be extended and orientated towards the object (i.e., around 70 cm). Additionally, we included an all-target condition to encompass a comprehensive range of interactions. By examining these variables, we aimed to understand better when and why the swarm manipulation technique might outperform or underperform relative to established techniques.

\begin{figure}[t]
  \centering
  \includegraphics[width=\linewidth]{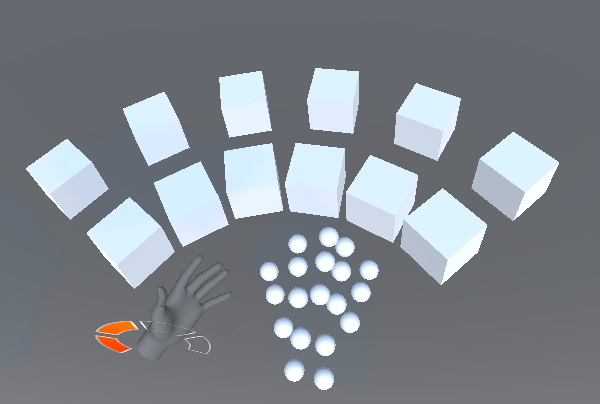}
  \caption{A screenshot of the Swarm Manipulation technique in VR. It showcases the non-dominant hand with a distribution bar encircling the left wrist, while the dominant hand interacts with the swarm. In the field of view, twelve targets are positioned ahead.}
  \label{fig:swarm}
\end{figure}

Dependent variables included: (a) \textit{Perceived Workload}, measured by raw NASA-TLX~\cite{hart_nasa-task_2006}, (b) \textit{Usability}, measured by System Usability Scale~\cite{brooke1996sus}, (c) \textit{User Experience}, measured by UEQ-Short~\cite{schrepp2017ueq}, and (d) \textit{Task Performance}, measured by task-completion time and deviations in rotation and resizing. Additionally, participants were asked to provide a final comparison and rate their preference and ease of use for the three manipulation techniques.

\subsubsection{Procedure}
After a brief introduction, participants were provided with a 5-minute period to familiarize themselves with the manipulation techniques and tasks that would be undertaken. To minimize the influence of learning effects, the order of manipulation techniques assigned to participants was counterbalanced using a Latin square design. All participants successfully completed the designated tasks utilizing each manipulation technique. The presentation of conditions within each assigned task was randomized, and each condition was repeated 10 times. Consequently, the overall study comprised a total of 5,400 trials, calculated as 3 (Technique) $\times$ 3 (Task) $\times$ 5 (Condition) $\times$ 10 (Repeat) $\times$ 12 (Participant).

After completing the tasks, participants were asked to fill out questionnaires evaluating their experience with the different manipulation techniques. Further, a final comparison was included, where participants were asked to rate their preference and the ease of use for each manipulation technique. Finally, a semi-structured interview was conducted to gather qualitative insights, allowing participants to share their strategies and provide suggestions. On average, participants spent approximately 30 minutes completing the study, including the tasks and questionnaire. Participants received a £5 reward for their participation. This user study was approved by our Department Ethics Application.

\subsubsection{Manipulation Techniques}
Our study employed the following interactions for each manipulation technique:

\textbf{\textit{Hand}}. Participants used Meta Quest hand tracking and could see their hand models in the VR environment. They touched the target object with their right hand for selection. In the Rotation task, the right hand was used to rotate the objects, with a grabbing gesture with the left hand confirming the action. The Resizing task followed the same procedure for selection, with scaling achieved by changing the distance between the right index fingertip and thumb tip.

\textbf{\textit{Controller}}. Participants used the Meta Quest~2 handheld controllers. The controller emitted a ray that could be manipulated in the VR environment. The selection was confirmed by pressing the pinch button on the controller when the ray pointed at the target object. The Rotation task involved the use of the ray and pinch buttons for object selection. The objects were rotated by rotating the controller, and the grip button was used for confirmation. The Resizing task followed a similar process for selection, with scaling achieved by dragging the ray outside or inside the objects.

\begin{figure}[t]
  \centering
  \includegraphics[width=\linewidth]{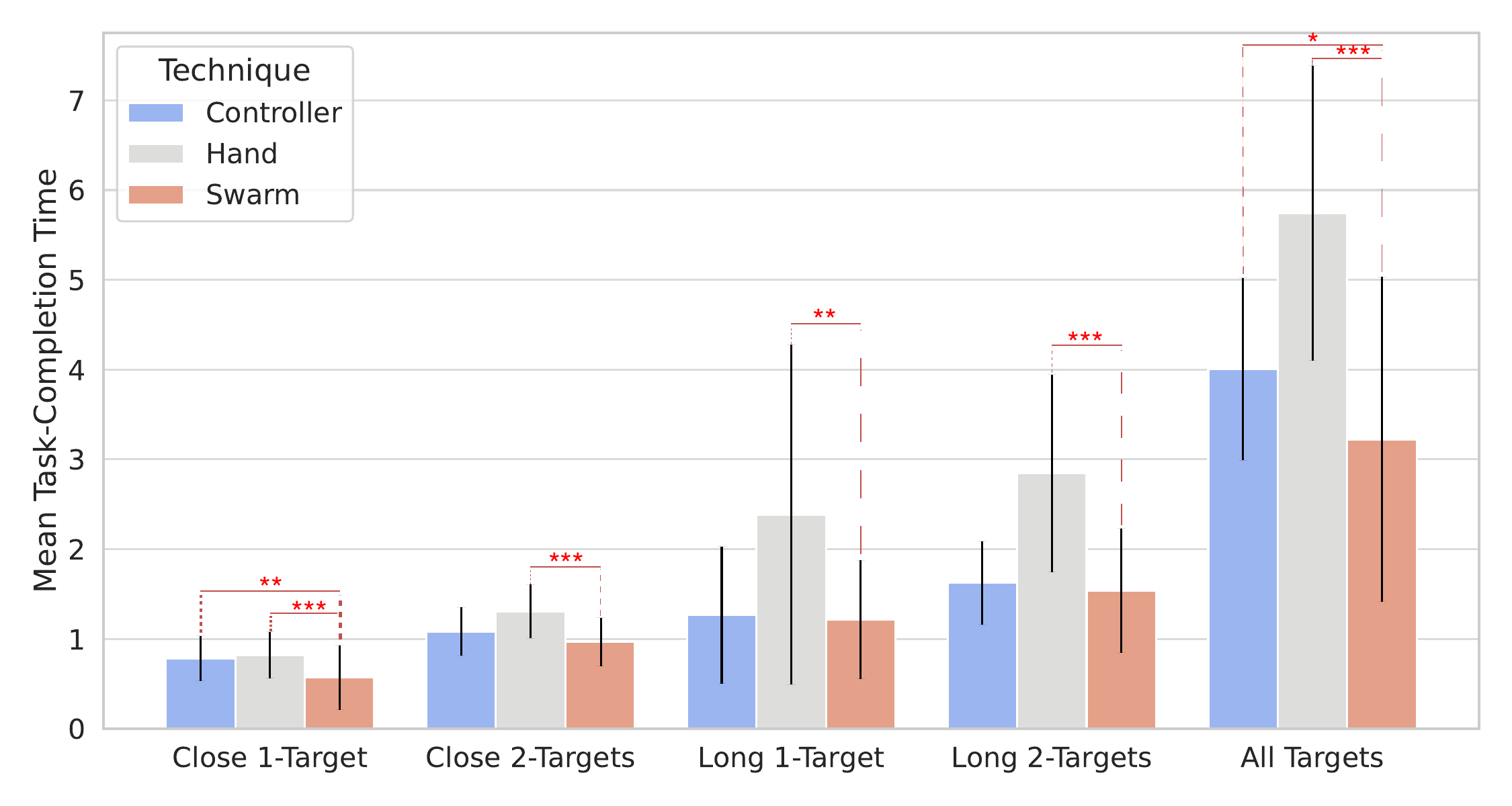}
  \caption{Mean task-completion times for each technique across five conditions of Task 1. Significant differences between conditions are annotated above the bars, with `*', `**', and `***' indicating significance levels at $p < 0.05$, $p < 0.01$, and $p < 0.001$, respectively.}
  \label{fig:task1}
\end{figure}

\textbf{\textit{Swarm Manipulation}}. Participants used the Meta Quest hand-tracking feature and could see their left-hand model, an indicator of Swarm Hand dispersion, and the Swarm Hand controlled by their right hand in the VR environment (see Figure \ref{fig:swarm}). The Selection task involved touching the target objects with any particle of the Swarm Hand (see Figure \ref{fig:teaser} (1)). The Rotation task followed the same procedure for selection, with object rotation achieved by rotating the right hand (see Figure \ref{fig:teaser} (2)). The Resizing task followed the same selection procedure, with scaling achieved by changing the distance between the right index fingertip and thumb tip (see Figure~\ref{fig:teaser} (3)).

\subsubsection{Tasks}

\textbf{\textit{Selection}}. Participants used the assigned manipulation technique to select the target objects, which were no longer highlighted in red upon selection.

\textbf{\textit{Rotation}}. Participants used the assigned manipulation technique to select target objects. A demonstration object appeared at the target angle (i.e., 45 degrees) once all target objects were selected. Participants were asked to rotate the selected target objects to match the demonstration object as closely as possible.

\textbf{\textit{Resizing}}. Participants used the assigned manipulation technique to select the target objects. A demonstration object appeared at the target size (1.2 times or 0.8 times the original object size) once all target objects were selected. Participants were asked to scale the selected target objects to match the demonstration object as closely as possible.

\subsubsection{Measures}

\textbf{\textit{Task Completion Time}}. The Task Completion Time in our study refers to the time taken by participants to complete various tasks, including selection, rotation, and resizing. This duration encompasses the entire process, starting from the generation of target objects for the selection task, continuing through the selection of target objects, and concluding with the participant's confirmation of task completion for each task.

\textit{\textbf{Rotation Angle Deviation:}} The deviation (°) between the final angle of the target object and the target angle set by the demonstration object in rotation tasks. This measure considered various rotation directions and angles chosen by the participants to achieve visually similar results.

\textit{\textbf{Resizing Size Deviation:}} The absolute deviation between the current size of the target object and the target size set by the demonstration object in resizing tasks.

\begin{figure}[t]
  \centering
  \includegraphics[width=\linewidth]{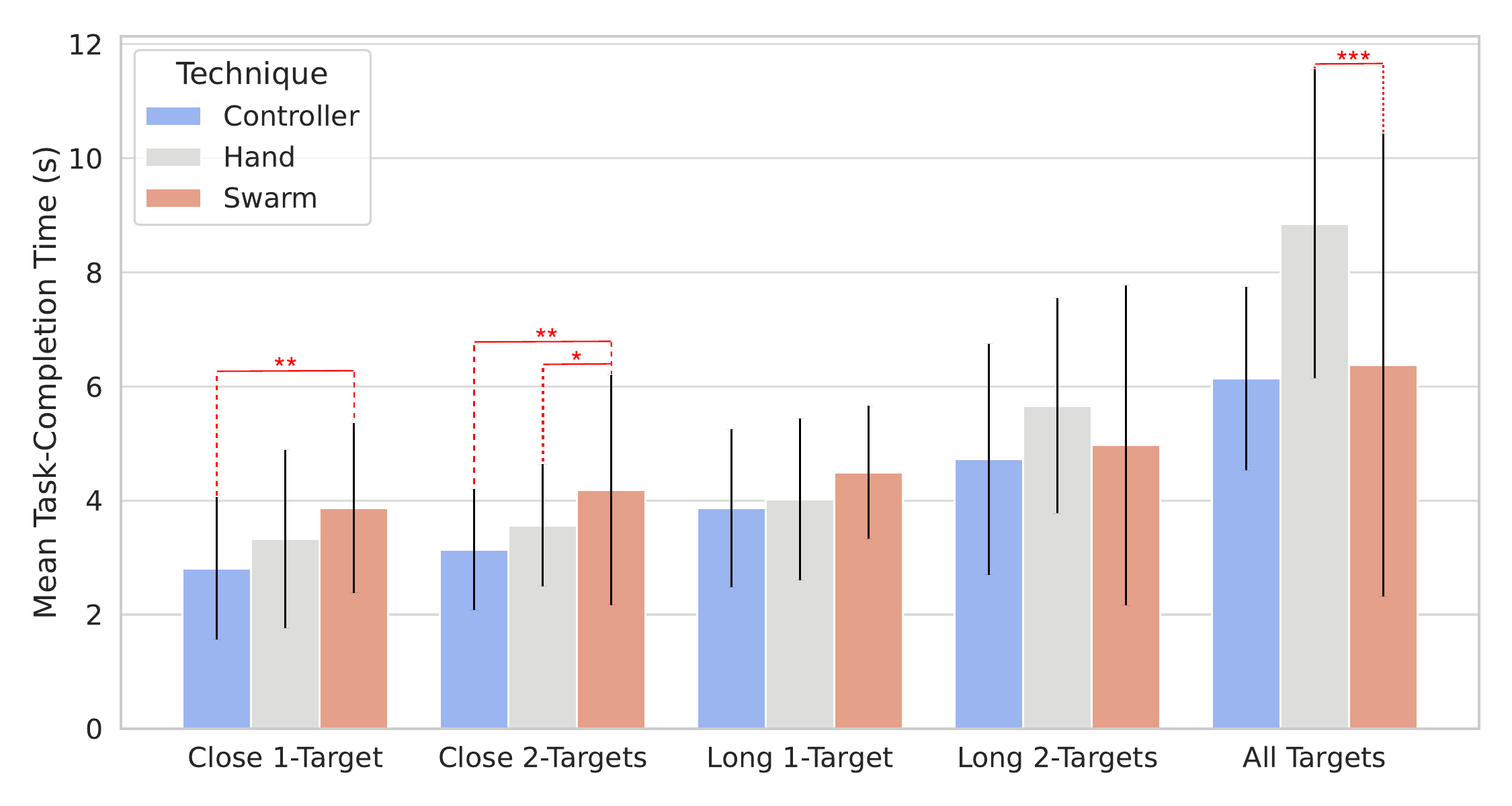}
  \caption{Mean task-completion times for each technique across five conditions of Task 2. Significant differences between conditions are annotated above the bars, with `*', `**', and `***' indicating significance levels at $p < 0.05$, $p < 0.01$, and $p < 0.001$, respectively.}
  \label{fig:task2_time}
\end{figure}

\begin{figure}[t]
  \centering
  \includegraphics[width=\linewidth]{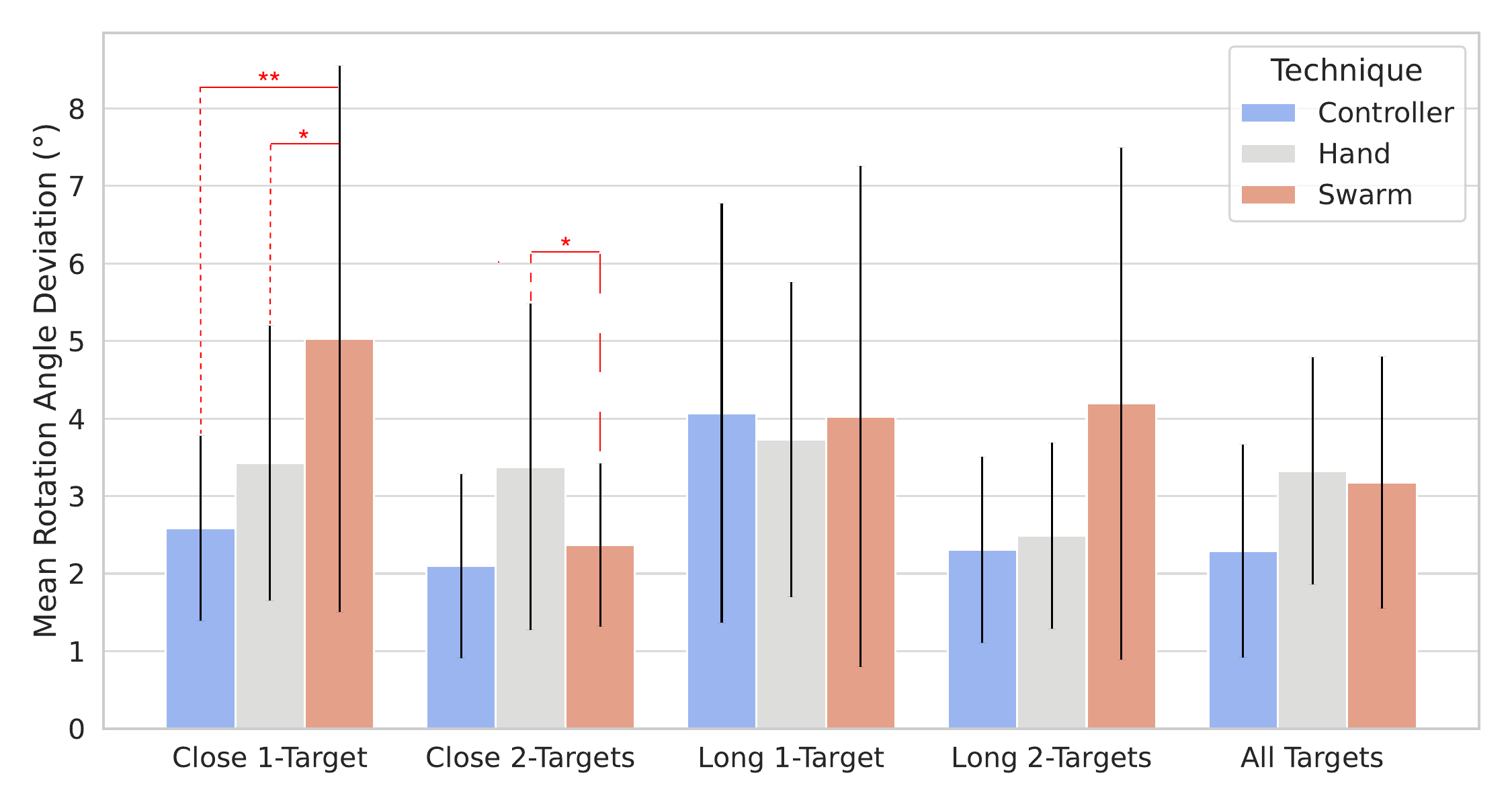}
  \caption{Mean rotation angle deviations for each technique across five conditions of Task 2. Significant differences between conditions are annotated above the bars, with `*' and `**', and `***' indicating significance levels at $p < 0.05$ and $p < 0.01$, respectively.}
  \label{fig:task2_de}
\end{figure}

\subsubsection{Participants and Apparatus}
A total of 12 participants (8 males and 4 females) were recruited for the study. The age range of the participants was between 20 and 23 years (\(M = 21.17, SD = .83\)). All participants were students at a local university. Most participants reported previous experience with VR, with familiarity ratings ranging from 0 to 6 on a 7-point Likert scale (\(M = 2.42, SD = 1.73\)), where 0 indicated no experience in VR, and 6 indicated expertise. All participants were right-handed habitual users.

The user study took place in a university laboratory equipped with a desktop computer, display devices, and an area for participants to engage in VR interactions using Meta Quest 2. The application used in the study was implemented in Unity 2021.3.23 and ran on a desktop computer.

\subsection{User Study 1: Results}

\textbf{\textit{Task 1: Task Completion Time}}. A two-way Analysis of Variance (ANOVA) was conducted to examine the main effects of technique and condition, as well as their interaction effect, on the task completion time. The main effect of the technique was significant with a partial Eta squared (\(\eta^2_p\)) of .067 (\(p < .001\)). The main effect of the condition was also significant with a \(\eta^2_p\) of .334 (\(p < .001\)). The interaction effect between the technique and condition was significant as well, with a \(\eta^2_p\) of .039 (\(p < .001\)).

\begin{figure}[t]
  \centering
  \includegraphics[width=\linewidth]{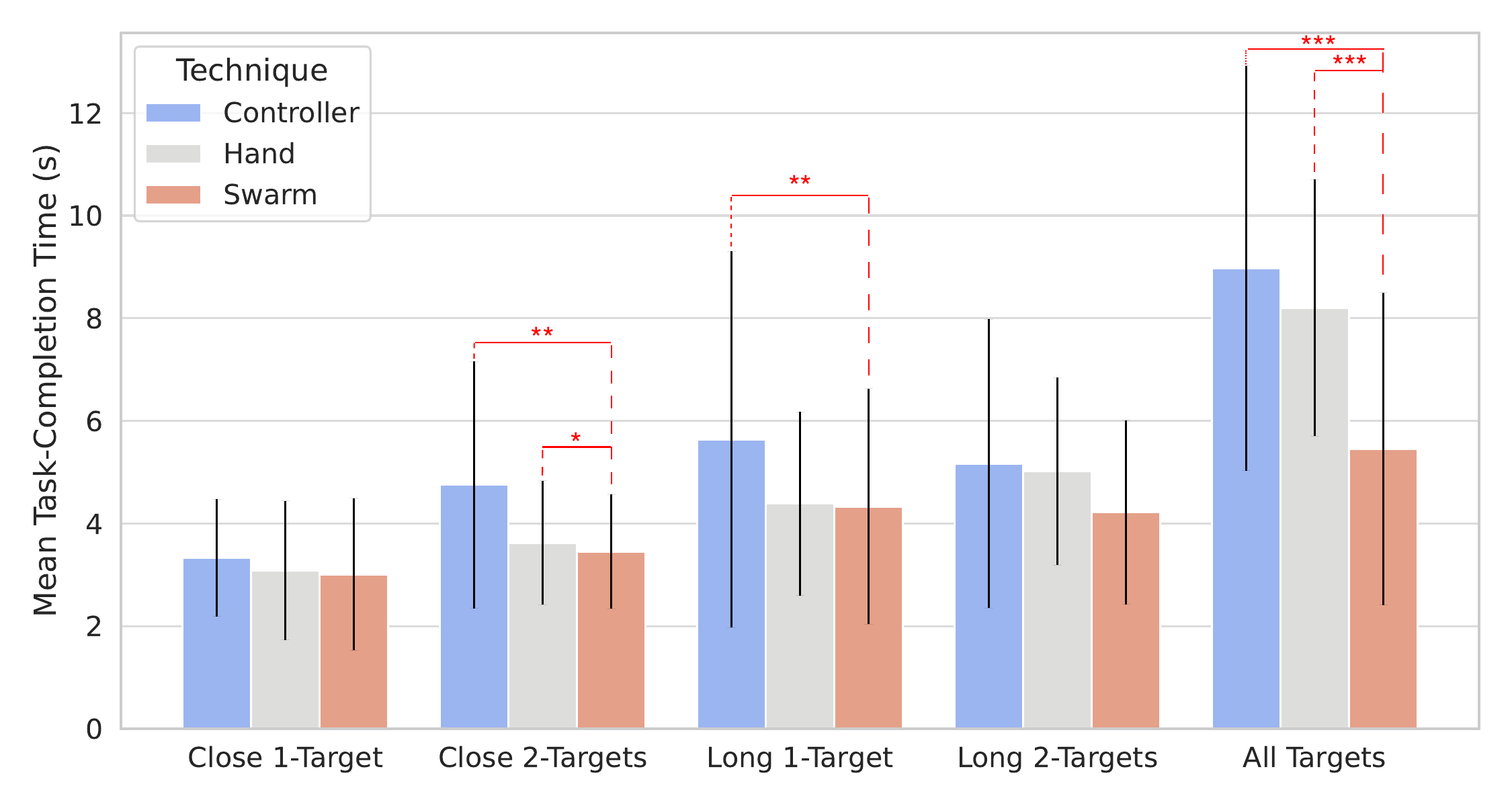}
  \caption{Mean task-completion times for each technique across five conditions of Task 3. Significant differences between conditions are annotated above the bars, with `*', `**', and `***' indicating significance levels at $p < 0.05$, $p < 0.01$, and $p < 0.001$, respectively.}
  \label{fig:task3_time}
\end{figure}

\begin{figure}[t]
  \centering
  \includegraphics[width=\linewidth]{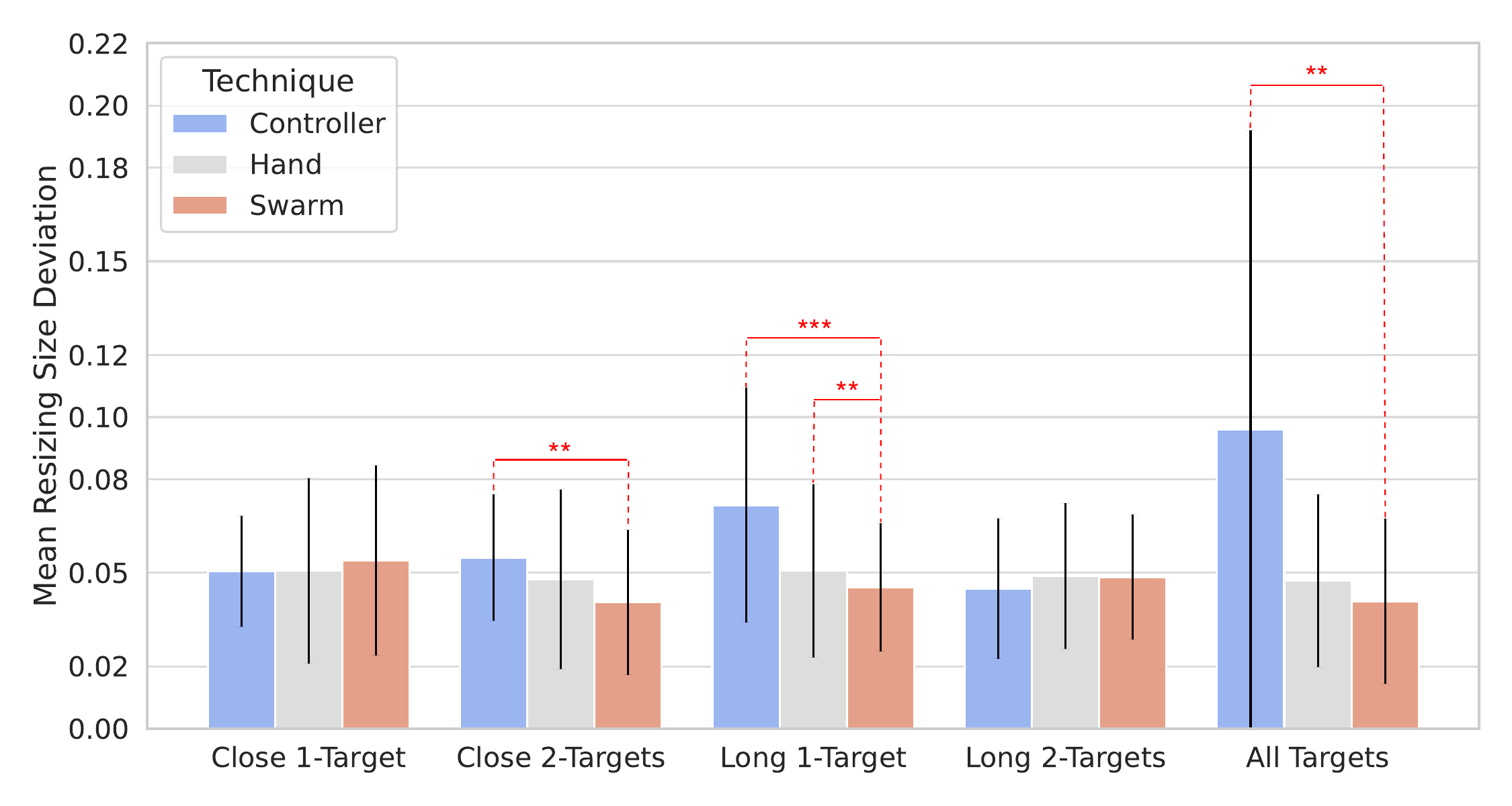}
  \caption{Mean resizing size deviations for each technique across five conditions of Task 3. Significant differences between conditions are annotated above the bars, with `**' and `***' indicating significance levels at $p < 0.01$ and $p < 0.001$, respectively.}
  \label{fig:task3_de}
\end{figure}

A post-hoc test using the Tukey HSD method was performed to make pairwise comparisons between the three techniques. In the close-distance single-target selection condition, Swarm was significantly faster than both Hand (\(p < .001\)) and Controller (\(p = .001\)). In the close-distance dual-target selection condition, Swarm was significantly faster than Hand (\(p < .001\)), but not significantly different from Controller (\(p = .148\)). In the long-distance single-target selection condition, Swarm was significantly faster than Hand (\(p = .006\)), but not significantly different from Controller (\(p = .900\)). In the long-distance dual-target selection condition, Swarm was significantly faster than Hand (\(p < .001\)), but not significantly different from Controller (\(p = .898\)). In all selection conditions, Swarm was significantly faster than both Hand (\(p < .001\)) and Controller (\(p = .016\)). The mean times for each technique across five conditions can be found in Figure \ref{fig:task1}.

\textbf{\textit{Task 2: Task Completion Time}}. A two-way ANOVA was conducted to examine the main effects of technique and condition, as well as their interaction effect, on the task completion time. The main effect of the technique was significant with a \(\eta^2_p\) of .022 (\(p < .001\)). The main effect of the condition was also significant with a \(\eta^2_p\) of .210 (\(p < .001\)). The interaction effect between the technique and condition was significant as well, with a \(\eta^2_p\) of .036 (\(p < .001\)).

A post-hoc test using the Tukey HSD method was performed to make pairwise comparisons between the three techniques. In the close-distance single-target rotation condition, Swarm was significantly slower than Controller (\(p = .001\)), but there was no significant difference between Swarm and Hand (\(p = .072\)). In the close-distance dual-target rotation condition, Swarm was significantly slower than both Hand (\(p = .044\)) and Controller (\(p = .001\)). There were no significant differences between Swarm and the other two techniques in the long-distance single-target rotation and long-distance dual-target rotation conditions. In the all-rotation condition, Swarm was significantly faster than Hand (\(p < .001\)), but there was no significant difference between Swarm and Controller. The mean times for each technique across five conditions of Task 2 can be found in Figure~\ref{fig:task2_time}.

\textbf{\textit{Task 2: Rotation Angle Deviation}}. A two-way ANOVA was conducted to examine the main effects of technique and condition and their interaction effect on the rotation angle deviation. The main effect of the technique was significant with a \(\eta^2_p\) of .011 (\(p < .001\)). The main effect of the condition was also significant with a \(\eta^2_p\) of .014 (\(p < .001\)). The interaction effect between the technique and condition was significant as well, with a \(\eta^2_p\) of .015 (\(p = .001\)).

\begin{figure*}[t]
  \centering
  \includegraphics[width=\linewidth]{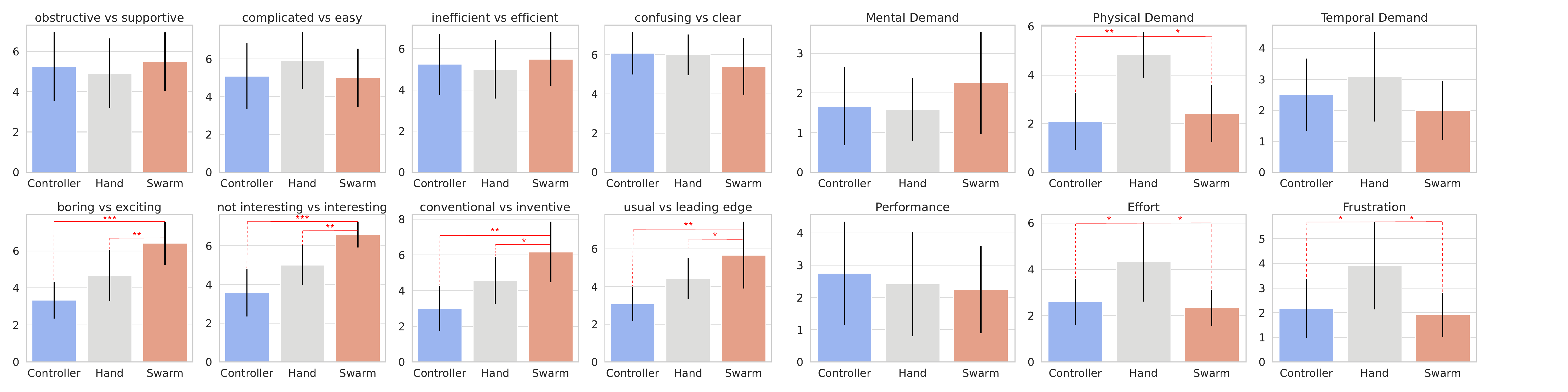}
  \caption{Bar charts illustrating the average ratings for eight UEQ subscales (left) and six NASA-TLX dimensions (right) across three techniques, with error bars representing standard deviation. Significant differences between conditions are annotated above the bars, with `*', `**', and `***' indicating significance levels at $p < 0.05$, $p < 0.01$, and $p < 0.001$, respectively.}
  \label{fig:ueq}
\end{figure*}

A post-hoc test using the Tukey HSD method was performed to make pairwise comparisons between the three techniques. In the close-distance single-target rotation condition, Swarm showed a significantly larger difference in rotation angle deviation than Controller (\(p = .001\)). There was also a significant difference between Swarm and Hand (\(p = .020\)). In the close-distance dual-target rotation condition, there was a significant difference between Swarm and Hand (\(p = .020\)), but no significant difference between Swarm and Controller. There were no significant differences between Swarm and the other two techniques in the long-distance single-target rotation and long-distance dual-target rotation conditions. In the all-rotation condition, there were no significant differences between Swarm and the other two techniques in terms of rotation angle deviation. The mean rotation angle deviations for each technique across five conditions of Task 2 can be found in Figure \ref{fig:task2_de}.

\textbf{\textit{Task 3: Task Completion Time}}. A two-way ANOVA was conducted to examine the main effects of technique and condition, as well as their interaction effect, on the task completion time. The main effect of the technique was significant with a \(\eta^2_p\) of .032 (\(p < .001\)). The main effect of the condition was also significant with a \(\eta^2_p\) of .164 (\(p < .001\)). The interaction effect between the technique and condition was significant as well, with a \(\eta^2_p\) of .024 (\(p < .001\)).

A post-hoc test using the Tukey HSD method was performed to make pairwise comparisons between the three techniques. In the close-distance single-target resizing condition, there were no significant differences in task completion times between Swarm and the other two techniques. In the close-distance dual-target resizing condition, Swarm was significantly faster than Controller (\(p = .001\)) and Hand (\(p = .044\)). In the long-distance single-target resizing condition, Swarm was significantly faster than Controller (\(p = .049\)), but there was no significant difference between Swarm and Hand (\(p = .072\)). There were no significant differences in task completion times between Swarm and the other two techniques in the long-distance dual-target resizing and all-resizing conditions. In the all-resizing condition, the Swarm Manipulation technique was significantly faster than both the Hand and Controller techniques (\(p < .001\)). The mean times for each technique across five conditions of Task 3 can be found in Figure \ref{fig:task3_time}.

\textbf{\textit{Task 3: Resizing Size Deviation}}. A two-way ANOVA was conducted to examine the main effects of technique and condition and their interaction effect on the resizing size deviation. The main effect of the technique was significant with a \(\eta^2_p\) of .008 (\(p < .001\)). The main effect of the condition was not significant (\(p = .175\)), and the interaction effect between the technique and condition was significant with a \(\eta^2_p\) of .012 (\(p = .0049\)). 

A post-hoc test using the Tukey HSD method was performed to make pairwise comparisons between the three techniques. In the close-distance single-target resizing condition, there were no significant differences between Swarm and the other two techniques. In the close-distance dual-target resizing condition, Swarm had significantly lower resizing size deviations than Controller (\(p = .007\)), but there was no significant difference between Swarm and Hand. In the long-distance single-target resizing condition, Swarm had significantly lower resizing size deviations than both Hands (\(p = .001\)) and Controller (\(p < .001\)). There were no significant differences between Swarm and the other two techniques in the long-distance dual-target resizing. In the all-resizing condition, Swarm had significantly lower resizing size deviations than Controller (\(p = .044\)). However, there was no significant difference between Swarm and Hand. The mean resizing size deviations for each technique across five conditions of Task 3 can be found in Figure~\ref{fig:task3_de}.

\subsubsection{Ratings and Preferences}

\textbf{\textit{User Experience Questionnaire}}. A Friedman test was conducted to compare each subscale of UEQ among the Controller, Hand, and Swarm techniques (see Figure \ref{fig:ueq} (left)). Significant differences were found in the following items: ``boring vs.~exciting'' ($\chi^2$(2) = 20.83, $p < .001$), ``not interesting vs.~interesting'' ($\chi^2$(2) = 15.83, $p < .001$), ``conventional vs.~inventive'' ($\chi^2$(2) = 14.22, $p < .001$), and ``usual vs.~leading edge'' ($\chi^2$(2) = 12.33, $p$ = .002). However, for the aspects ``obstructive vs.~supportive'' ($\chi^2$(2) = .35, $p$ = .839), ``complicated vs.~easy'' ($\chi^2$(2) = 1.95, $p$ = .378), ``inefficient vs.~efficient'' ($\chi^2$(2) = 1.72, $p$ = .423), and ``confusing vs.~clear'' ($\chi^2$(2) = 3.19, $p$ = .203), there were no significant differences between the techniques. A post-hoc analysis was conducted using a pairwise Wilcoxon Rank-Sum Test (Mann-Whitney U test) with Bonferroni correction for multiple comparisons. The post-hoc analysis showed significant differences in most comparisons, except for ``not interesting vs. interesting'' between Controller and Hand techniques ($W = 5, p = .052$), and ``usual vs. leading edge'' between Hand and Swarm techniques ($W = 7, p = .103$).


We also calculated the scores for Hedonic Quality, Pragmatic Quality, and Overall scores for UEQ. Significant differences were found in Hedonic Quality (\(\chi^2\)(2) = 17.91, \(p < .001\)) and Overall scores (\(\chi^2\)(2) = 8.98, \(p = .0112\)). However, for Pragmatic Quality, there were no significant differences between the techniques (\(\chi^2\)(2) = 1.38, \(p = .502\)). The post-hoc analysis showed significant differences for Hedonic Quality between Controller and Hand (\(W = 0.75\), \(p = .0057\)), Controller and Swarm (\(W = 0.95\), \(p = .0002\)), and Hand and Swarm (\(W = 0.74\), \(p = .0062\)). For Overall, significant differences were found between Controller and Swarm (\(W = 0.82\), \(p = .0021\)). No significant differences were found for Overall between Controller and Hand (\(W = 0.47\), \(p = .157\)) and Hand and Swarm (\(W = 0.53\), \(p = .0893\)).

\begin{table}[ht]
\centering
\begin{tabularx}{\linewidth}{|l|X|X|X|}
\hline
\textbf{Technique} & \textbf{Pragmatic Quality} & \textbf{Hedonic Quality} & \textbf{Overall} \\
\hline
Controller  & 1.42 (1.15) & -0.75 (0.88) & 0.34 (0.67) \\
Hand        & 1.46 (0.90) & 0.67 (0.94) & 1.06 (0.81) \\
Swarm       & 1.35 (1.33) & 2.21 (0.98) & 1.78 (0.80) \\
\hline
\end{tabularx}
\caption{Mean and standard deviation (SD) for Pragmatic Quality, Hedonic Quality, and Overall for each technique.}
\label{table:summary_stats}
\end{table}

\textbf{\textit{NASA Task Load Index}}. A Friedman test was conducted to compare the effect of different techniques on six measures of the NASA Task Load Index: Mental Demand, Physical Demand, Temporal Demand, Performance, Effort, and Frustration (see Figure \ref{fig:ueq} (right)). The results showed significant differences in ``Physical Demand'' (\(\chi^2(2) = 15.24, p < .001\)), ``Effort'' (\(\chi^2(2) = 11.35, p = .0034\)), and ``Frustration'' (\(\chi^2(2) = 11.53, p = .0031\)) across the techniques. However, no significant differences were found in ``Mental Demand'' (\(\chi^2(2) = 4.33, p = .115\)), ``Temporal Demand'' (\(\chi^2(2) = 5.25, p = .072\)), or ``Performance'' (\(\chi^2(2) = 2.11, p = .347\)) across the techniques.

A post-hoc analysis was conducted using pairwise Wilcoxon Rank-Sum Tests (Mann-Whitney U tests) with Bonferroni correction for multiple comparisons. The post-hoc analysis revealed significant differences in the ``Physical Demand'' aspect. Specifically, for ``Physical Demand,'' the Hand technique exhibited significantly higher scores compared to both the Swarm and Controller techniques ($W = 1.0, p = .003$), and the Hand technique also had significantly higher scores compared to the Swarm technique ($W = .0, p = .010$). Regarding ``Effort,'' we observed significant differences between the Controller and Hand techniques. The Effort scores were significantly higher for the Hand technique compared to the Controller technique ($W = 2.5, p = .018$), and similarly, the Hand technique had significantly higher scores compared to the Swarm technique ($W = 3.0, p = .021$). For the ``Frustration'' aspect, significant differences were found between the Controller and Hand techniques. The Frustration scores were significantly higher for the Hand technique compared to the Controller technique ($W = 5.5, p = .041$), and also significantly higher for the Hand technique compared to the Swarm technique ($W = 2.0, p = .026$).


\textbf{\textit{System Usability Scale}}. A Friedman test was conducted to compare the SUS scores among the Controller ($M = 87.08, SD = 14.18$), Hand ($M = 83.96, SD = 13.75$), and Swarm techniques ($M = 81.25, SD = 14.16$). The results revealed no significant difference in SUS scores across the three techniques (\(\chi^2(2) = 1.22, p = .544\)). These results suggest that there is no significant difference in usability, as measured by the SUS, between the three techniques. However, as suggested by \cite{brooke1996sus}, a SUS score of 70 means the system usability is at an acceptable level. This indicates that the participants have been very receptive to the system usability of Swarm Manipulation, especially given that the other two conditions are already widely used commercially in VR.

\textbf{\textit{Preference}}. Regarding the performance ranking, eight participants ranked Swarm Manipulation as the best technique, while Controller was ranked best by three participants, and Hand was only ranked as the best technique by one participant. In terms of ease of use, Swarm Manipulation received the highest average rating of 4.5 out of 5 ($SD = .90$) among the three techniques. The Controller was rated 3.83 ($SD = .94$), while Hand received an average rating of 3.75 ($SD = .97$).

\subsection{User Study 1: Discussion}
\subsubsection{Task Performance}
The task performance results shed light on the effectiveness and potential advantages of the Swarm Manipulation technique compared to the Hand and Controller techniques in different VR manipulation tasks. In the selection task, the Swarm Manipulation technique demonstrated significantly faster completion times compared to the Hand technique in various conditions and mixed results compared to the Controller. This suggests that the swarm-based approach can enhance efficiency and speed in target selection, though the advantages over the Controller may vary depending on the conditions.

In the rotation task, the Swarm Manipulation technique outperformed the Hand technique in terms of task completion time only in the all-target rotation condition and was significantly slower in both close-distance dual-target conditions. This finding highlights the complexity of the rotation task and suggests that the Swarm Manipulation technique's efficiency may be context-dependent. The lower rotation angle deviation in the close-distance dual-target condition indicates a trade-off between speed and accuracy, possibly due to the distributed and collective nature of swarm entities.

In the resizing task, the Swarm Manipulation technique was significantly faster than the Controller in many conditions but was only significantly faster than the Hand in the all-target resizing condition. However, the Swarm manipulation approach did demonstrate significantly smaller resizing size deviations compared to Hand in the long-distance single-target resizing condition, implying higher accuracy in certain scenarios. The Swarm Manipulation technique also demonstrated significantly smaller resizing size deviations than Controller in the close-distance dual-target condition, long-distance single-target condition, and all-target resizing condition. It is essential to recognize the outstanding performance of the Swarm Manipulation technique in resizing tasks, as it excelled in terms of speed and accuracy under different conditions.

Overall, our Swarm Manipulation technique shows promising advantages in speed and accuracy in certain tasks and conditions. Still, the performance is not uniformly superior across all scenarios, especially in the Rotation task. Further investigation and refinement could focus on understanding the underlying factors that contribute to these variations in performance. The collective control theory and coordination of swarm entities might help compensate for the potential challenges introduced by distance, resulting in more accurate resizing manipulations.

\begin{figure*}[t]
  \centering
  \includegraphics[width=\linewidth]{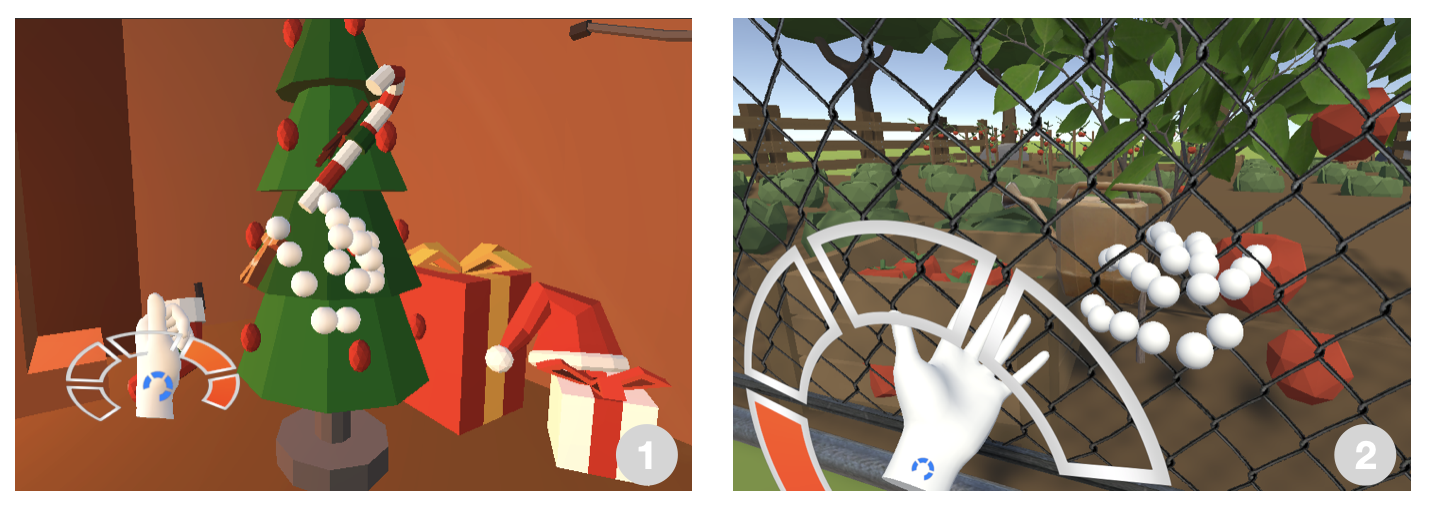}
  \caption{Two open-ended scenarios demonstrating the application of the Swarm Manipulation technique. (1) Left: Participants engage in a task involving the selection and manipulation of various objects, some obscured by a Christmas tree, challenging them to adapt their manipulation strategies; (2) Right: the scenario requires participants to navigate through grid obstacles using the swarm manipulation technique, facilitating an exploration of body ownership and perceptual changes in a swarm state.}
  \label{fig:user_study2}
\end{figure*}

\subsubsection{User Preference}
Based on our analysis of the user preference questionnaires we gathered, it is evident that the Swarm Manipulation technique was generally well-received by the participants. The majority of participants not only ranked it as the most effective technique ($N = 9$) but also considered it the most user-friendly ($N = 10$).

Nevertheless, it is important to acknowledge the presence of diverse opinions among the participants. For instance, \textit{P1} commented that the Swarm Manipulation technique provided an expanded range of finger control, enabling the manipulation of objects without the need for repetitive hand movements. However, this user also pointed out certain difficulties with gesture recognition, indicating room for improvement in terms of accuracy. Conversely, \textit{P2} found the Swarm Manipulation technique innovative and valued its potential for adjusting reach length. Yet, they also emphasized the necessity for improved gesture recognition and stability due to issues encountered with the system misinterpreting their gestures. Additionally, \textit{P11} suggested that incorporating a visual representation of the hand's outline within the Swarm Manipulation technique could be advantageous for novice VR users, aiding their orientation in the virtual environment. \textit{P1}, \textit{P2}, and \textit{P10} also emphasized the challenge of learning the manipulation mechanisms, highlighting the need for enhanced user guidance and support.

While the overall reception of the Swarm Manipulation technique was positive, the feedback received clearly indicates areas in need of improvement, namely gesture recognition accuracy, system stability, and user guidance. Further research endeavours could explore these specific areas to explore them more comprehensively, ultimately enhancing the usability and overall user experience of the Swarm Manipulation technique in the context of VR.

\section{User Study 2: Qualitatively Investigate Two Case Studies of Deploying Swarm Manipulation in VR}

Given that our first user study was very focused on comparing the efficiency and differences between swarm hands and the other two manipulation techniques in a neatly arranged VR scene, our Study 2 explores how users employ swarm interactions in more realistic and open-ended scenarios in VR, focusing on the transformation of user ownership, agency, and perception through the use of the Swarm Manipulation technique. Therefore, we conducted a second user study, and this time, we gathered only a minimum of Likert scale data and subjective discussions from a small group of new participants \cite{tao2022integrating}, as we were mostly interested in their open-ended feedback to guide the development of our future work.

\subsection{User Study 2: Method}
\subsubsection{Study Design}
Study 2 is exploratory, emphasizing subjective feedback and minimal quantitative data. Participants engaged in two open-ended scenarios using the Swarm Manipulation technique to understand user strategies and perceptions in more naturalistic settings. The two scenarios were selected to explore the Swarm Manipulation technique in more realistic and complex VR settings. The first scenario involved object selection and manipulation with occlusions, which mimics real-world challenges where objects are not always fully visible. The second scenario required participants to navigate through grid obstacles, emphasizing the flexibility and adaptability of the swarm interaction technique in constrained environments. These scenarios required different interaction strategies compared to the more controlled tasks in the first study, providing deeper insights into the practical applications of swarm manipulation in VR.

\subsubsection{Procedure}

Participants began with a 5-minute training session on the Swarm Manipulation technique, with an emphasis on wrist rotation to adjust the distribution of swarm particles for better control. After completing the tasks, participants were asked to fill out questionnaires to evaluate their experience with the different manipulation techniques. Subsequently, a semi-structured interview was conducted to gather qualitative insights, allowing participants to provide suggestions for optimizing the control of swarm particles and sharing their insights on how their perception changed while using a swarm-like hand in VR. Participants spent approximately 20 minutes completing the questionnaire. After all, participants received a £5 reward for their participation. This user study shared the same approval from our Department Ethics Application.

\subsubsection{Tasks}

In the first scenario, as depicted in Figure \ref{fig:user_study2} (1), participants engaged in object selection and manipulation within an environment featuring a Christmas tree. This setting necessitated the adaptation of swarm particle distribution techniques to facilitate interactions with objects that were partially occluded.

In the second scenario, illustrated in Figure \ref{fig:user_study2} (2), participants were tasked with navigating obstacles (grilles) through the utilization of swarm-like body interactions. This task aimed to investigate participants' perception of body ownership and their experiences within swarm-state interactions.

For both scenarios, no specific objectives were provided to the participants. Instead, they were encouraged to explore all facets of the Swarm Manipulation technique, including the manipulation of swarm particle distribution through wrist rotation and the comparison of swarm-like body experiences with the conventional continuous intact body paradigm. The primary goal was to encourage participants to interact with as many objects as possible within the virtual environment while engaging with the unique aspects of the Swarm Manipulation technique.

\subsubsection{Measures}
\textbf{\textit{Likert Scale Questionnaire}}. Participants responded to three questions on a 7-point Likert scale (ranging from 0 to 6), addressing aspects of swarm particle control, enhancement of immersion, and a sense of bodily control in the virtual environment.

\begin{enumerate}
    \item [Q1:] \textit{When selecting objects under occluded conditions, to what extent did you tend to adjust the distribution to control the swarm particle's arrangement and distribution for grabbing objects?}
    \item [Q2:] \textit{How much did using the Swarm Manipulation technique enhance your sense of immersion in the virtual environment?}
    \item [Q3:] \textit{How much did using the Swarm Manipulation technique enhance a sense of control in the virtual environment that you do not have in real life?}
\end{enumerate}

\textbf{\textit{Open-ended Comments}}. Participants also provided detailed feedback based on their Likert scale responses, focusing on their experiences, comparisons with traditional VR interactions, and suggestions for further improvements.

\subsubsection{Participants and Apparatus}
A total of 6 participants (4 males and 2 females) were recruited for the study from a local university. The age range of the participants was between 20 and 22 years (\(M = 21.5, SD = .5\)). All participants reported previous experience with VR, with familiarity ratings ranging from 0 to 6 on a 7-point Likert scale, where 0 indicated no experience in VR, and 6 indicated expertise (\(M = 2.33, SD = 1.14\)). None of the participants took part in User Study 1. This study followed the same apparatus as Study 1.

\subsection{User Study 2: Results}

\textbf{\textit{Adjusting Swarm Particle Distribution}}
Participants showed a diverse range of tendencies in adjusting the swarm particle distribution, with the mean response being 3.67 ($SD = 1.63$). This variation suggests that while some participants found it relatively easy and intuitive to adjust the swarm particles (as indicated by higher scores), others were less inclined to do so, preferring alternative methods or facing challenges in the adjustment process.

\textbf{\textit{Enhancement of Immersion}}
The responses to the second question, concerning the enhancement of immersion in the virtual environment using the Swarm Manipulation technique, showed a higher level of agreement among participants. The mean score was 4.50 ($SD = 1.38$), indicating that most participants felt a significant increase in immersion when using this technique.

\textbf{\textit{Enhancement of Control}}
The responses for the enhancement of a sense of control in the virtual environment were consistently high, with a mean score of 5.17 ($SD = .41$). This indicates a strong consensus among participants that the Swarm Manipulation technique provided an enhanced sense of control that is distinct from real-life control, underscoring this technique's potential in creating unique VR interaction experiences.

\subsection{User Study 2: Qualitative Findings}

\textbf{\textit{Interacting with Swarm Particles}}. One key takeaway from our participants is the intuitive and natural feel of the Swarm Manipulation technique. \textit{P1} noted, \textit{``This method is easy to operate and provides a refreshing sensory experience,''} suggesting that the technique offers a unique and engaging interaction experience. The fluidity and flexibility of the swarm particle body, as described by \textit{P1}, hint at the potential for providing users with a more immersive and intuitive interaction experience in mixed reality.

Participants also highlighted the importance of adaptability in controlling the distribution of swarm particles. \textit{P6} emphasized that the technique allows for \textit{``flexible adjustment to adapt to the current environment,''} indicating the advantage of adjusting the dispersion of swarm particles to suit different tasks and scenarios. This adaptability aligns well with the dynamic nature of mixed reality environments. Further research into fine-tuning these control mechanisms is warranted to optimise the control of swarm interaction to ensure they remain responsive and intuitive across diverse scenarios.

Furthermore, the participants' comments emphasized the significance of user habits developed in the physical world. \textit{P2} pointed out that in reality, people tend to \textit{``extend their arms rather than their palms''} when grabbing objects. This insight suggests that addressing ingrained habits through user training and guidance may be necessary to optimize swarm interaction in mixed reality. Bridging the gap between physical and virtual behaviours, as suggested by \textit{P3}, could further enhance the user experience and facilitate a smoother transition to swarm-based interactions.

The concept of controlling the shape of the swarm particles was highlighted as an advantage by \textit{P4}, who noted that it allows users to \textit{``change the shape of the hands and freely pass through gaps.''} This dynamic control offers potential benefits in terms of adaptability and precision in object manipulation. However, it also poses design challenges in maintaining a balance between user agency and ease of control. Optimizing this aspect of swarm interaction will require careful consideration of user preferences and task requirements.

Another noteworthy insight was \textit{P5}'s suggestion to focus on minimal movements and fingertip actions when controlling swarm particles. They mentioned that \textit{``control actions with the swarm particles should be as small as possible, primarily relying on fingertip actions.''} This aligns with the broader trend in mixed reality and VR interfaces, emphasizing natural and effortless interactions. To optimize swarm interaction, it is essential to refine the hand gestures and control mechanisms to ensure they are both intuitive and ergonomic, minimizing user fatigue.

\textbf{\textit{Enhancement of Immersive Experience}}. The enhancement of immersion was a prominently acknowledged benefit of swarm interaction, as evidenced by the responses to Q2. \textit{P3} particularly underscored this aspect, stating, ``The synchronization of movement and visuals, along with the realism of virtual objects, significantly impacts the sense of immersion.'' This observation is in agreement with the high mean scores for Q2, underscoring the importance of synchrony and realistic visual feedback in VR to create deeply immersive experiences.

Besides, as highlighted by \textit{P2}, customization and adaptability play a vital role in user engagement and immersion. Allowing users to adjust the distribution of swarm particles provides a sense of control expansion, contributing to a more immersive experience. To optimize swarm interaction, developers should provide customization options, enabling users to tailor swarm behavior to their preferences, thereby increasing user agency and satisfaction.

The perception of the swarm form, as discussed by \textit{P5}, underscores the importance of balancing abstraction and familiarity. While the swarm form may feel more abstract, familiarity with scenarios can aid user understanding and task focus. Striking this balance can optimize user immersion, ensuring a sense of connection with the virtual environment.

\textbf{\textit{Towards Understanding Swarm Control}}. A critical aspect of this study was the enhanced sense of control experienced by users in the VR environment, a feature that stood out in the responses to Q3. As \textit{P6} articulated, ``The discrete interactions of Swarm Hand in VR contrast with continuous real-life interactions, offering a new and exciting experience.'' This sentiment reflects the high scores recorded for Q3 and highlights the unique opportunities presented by the Swarm Manipulation technique in crafting VR experiences that are distinctively different from real-world interactions. As \textit{P2} noted, it introduces a novel approach to control by enabling users to manipulate the distribution of swarm particles. This novelty suggests that future developments in swarm interaction should continue to explore innovative ways to provide users with fresh and engaging control mechanisms, contributing to a heightened sense of control and immersion.

Additionally, \textit{P3} highlighted the potential advantages of the Swarm Manipulation technique for enhancing hand coordination and flexibility. The transition from continuous and intuitive real-world interactions to discrete and innovative virtual interactions offers users a new and valuable experience. This insight suggests that swarm interaction can be optimized by designing experiences that encourage users to develop and improve their coordination and adaptability in the virtual environment.

\begin{figure*}[t]
  \centering
  \includegraphics[width=\linewidth]{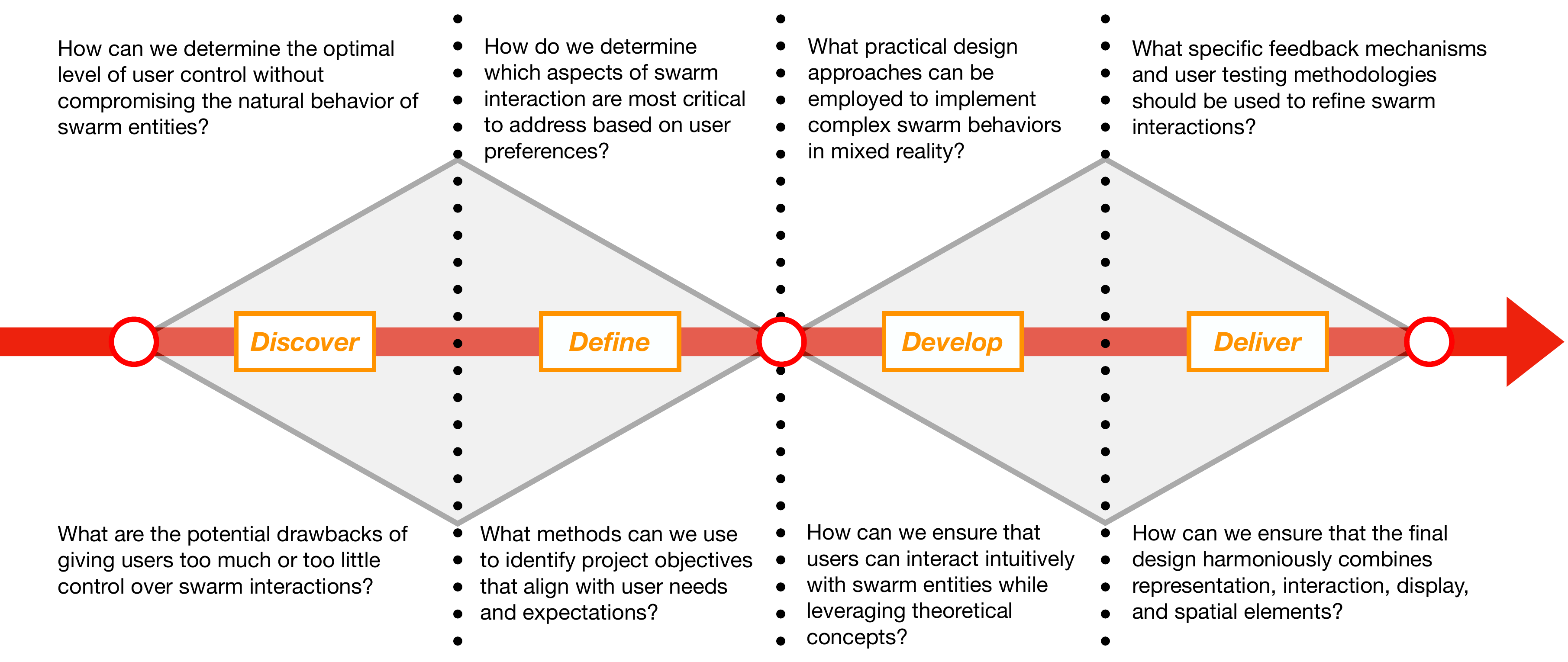}
  \caption{The Double Diamond design process \cite{council2005double} for swarm interaction in mixed reality. This diagram illustrates the iterative design phases—--Discover, Define, Develop, and Deliver—---along with their respective challenges for creating swarm interactions in mixed reality.}
  \label{double_diamond}
\end{figure*}

Another significant point raised by \textit{P5} is that swarm interaction allows for finer and more nuanced control, potentially differing from interactions in the physical world governed by strict physical laws. This aspect not only enhances users' sense of control but also showcases the unique possibilities of mixed reality environments. Future optimization efforts should focus on leveraging swarm interaction to enable precise and subtle actions that go beyond the limitations of the physical world.

However, the study also identified areas for improvement in the Swarm Manipulation technique. \textit{P3, P5, and P6} suggested enhancements such as more dispersed and visible particle distribution and more intuitive control mechanisms. These insights are invaluable for guiding future developments in the technology. Additionally, the study emphasizes the critical role of realistic synchronization and visual feedback in enhancing the user experience in VR, pointing towards essential aspects to be considered in future iterations of VR technology design.

\section{Discussions, Limitations, and Future Work}

This paper represents an initial exploration and does not attempt to cover the entire spectrum of factors influencing the manipulation of swarm entities in VR environments. We acknowledge the limited number of participants in our first study ($N$ = 12) and the second study ($N$ = 6) as a constraint. However, the second study aimed to extend our understanding of swarm manipulation techniques through qualitative insights in more realistic scenarios rather than focusing solely on quantitative metrics such as accuracy and speed. The semi-structured interviews conducted with the participants provided valuable feedback on the usability and potential improvements of the Swarm Manipulation technique. Our forthcoming endeavors will be directed towards further investigations along this trajectory. A promising starting point is a comprehensive analysis of the control models~\cite{Abtahi_19,Abtahi_22,Gonzalez_Sensorimotor} underpinning swarm interactions.

Additionally, the primary emphasis of this paper lies in the evaluation of interaction tasks using a prototype of this novel interaction technology, with limited exploration of the underlying interaction theory~\cite{hornbaek2017interaction}. In our future work, we plan to employ computational interaction methods~\cite{oulasvirta2018computational} to gain insights into users' understanding of swarm interaction and optimise the associated control model, aiming to further enhance performance. However, it is worth highlighting that despite our limited understanding of the control principles governing Swarm Manipulation, our research has already demonstrated significant benefits in manipulation tasks, thereby underscoring the potential of this field to define new interaction paradigms in the future~\cite{mueller2021,mueller2023}. 

Following the methodology outlined by Le Goc et al. \cite{le_goc_zooids_2016}, we present our design space on swarm interactions in virtual reality, which is compartmentalized into specific aspects, each containing multiple dimensions. This categorization serves as a foundational schema for investigating and developing swarm interaction techniques in mixed reality, enabling exploration and novel contributions. To use the design space in the design of swarm interaction for MR, we suggest leveraging the double diamond design process \cite{council2005double}, see Figure~\ref{double_diamond} (adapted from the Double Diamond design process \cite{council2005double}, which offers a visual depiction of the iterative design phases and associated challenges). For each of the four phases---discover, define, develop, and deliver---we suggest using the challenges outlined in Figure~\ref{double_diamond} as prompts to revisit the aspects and their individual dimensions as outlined in the design space. In this way, a design or research team can ensure all relevant aspects of swarm interaction in mixed reality are taken into account when diverging and convergent through the design process.

Moreover, we also outline a vision for our future research. We firmly believe that in addition to investigating how users control swarm interaction, an essential topic for subsequent studies, exploring how swarm particles \emph{collaborate} and exchange signals among themselves (that is, ``self-organising''), and consequently applying automation theory, is central to advancing research on swarm interaction. This represents a novel interpretation of Pattie Maes' vision of Intelligent Agents~\cite{maes_95,maes1995agents}, while retaining the concept of Direct Manipulation~\cite{shneiderman1997direct}.

Finally, these swarm particles may serve as the fundamental building blocks of future VR/AR interaction interfaces. Similar to Hiroshi Ishii's Tangible Bits~\cite{ishii_bits} in the realm of tangible user interfaces and concepts such as Radical Atoms~\cite{ishii_atoms} in the future materials field, swarm particles can be regarded as the smallest units for user control and interaction, as well as for constructing virtual worlds. The advantage of using swarm particles for display and interaction is that they resemble pixel dots in the digital world, are less expensive to maintain than tangible user interfaces~\cite{Lars_2023}, and allow developers to design adaptive user interfaces that balance visuals with immersive experiences. Consequently, swarm particles have the potential to become the elemental units that aid researchers in understanding how to interact with virtual information in the future.

\section{Conclusion}

In summary, we introduced the Swarm Manipulation technique for VR interactions and performed a thorough evaluation, comparing its performance against traditional Hand and Controller manipulation techniques, with the participation of 12 individuals. Our findings offer a nuanced perspective on the subject matter. Additionally, we explored open-ended interactions with 6 participants across two complex VR scenarios, collecting their subjective feedback to guide the optimization of our future steps in swarm interactions.

Overall, our results demonstrate that the Swarm Manipulation technique exhibits excellent performance, showcasing significantly faster speeds compared to at least one of the other two techniques across most tested conditions for various tasks. Notably, the Swarm technique's advantage in rotation tasks is context-dependent, showing increased speed in close-distance single-target scenarios while also highlighting a trade-off between speed and accuracy, attributed to the distributed and collaborative nature of swarm particles. In resizing tasks, the performance of the Swarm Manipulation technique varied, with increased speed in specific conditions and minimized size deviations in the long-distance single-target resizing scenario. However, these advantages were not consistently observed across all resizing conditions.

Furthermore, the subjective user feedback gathered from participants unequivocally positions the Swarm Manipulation technique as the most preferred and user-friendly approach among the manipulation techniques under scrutiny. This feedback underscores the potential of the Swarm Manipulation technique as a promising method for VR object manipulation. Our paper serves as an initial proof-of-concept, revealing its usability enhancement in specific scenarios while highlighting areas where performance variation necessitates further investigation.

Additionally, our second user study explores more realistic and open-ended VR scenarios, exploring the transformation of user ownership, agency, and perception through the use of the Swarm Manipulation technique. We also introduce a Double Diamond design process to guide future research on designing swarm interaction techniques in mixed reality.

Our future research endeavours will investigate deeper into the realm of swarm interactions in mixed reality, focusing on understanding user control dynamics and collaborative particle behaviour within the swarm paradigm. Building upon the complex performance patterns identified in this paper, we aim to refine and optimize the Swarm Manipulation technique for a broader range of applications and conditions.

\section*{Acknowledgments}
Xiang Li is supported by the China Scholarship Council (CSC) International Cambridge Scholarship (No. 202208320092). John J. Dudley and Per Ola Kristensson are supported by the EPSRC (grants EP/S027432/1 and EP/W02456X/1).

\bibliographystyle{cag-num-names}
\bibliography{refs}


\end{document}